# Effect of pseudo-cubic (111)-oriented orthorhombic substrate facets on perovskite oxide thin film synthesis

K. Kjærnes[1], T. Bolstad[1], D. M. Evans[2], E. Lysne[2], B. A. D. Williamson[2], D. Meier[2], S. M. Selbach[2], T. Tybell[1, *]

[1] Department of Electronic Systems, NTNU - Norwegian University of Science and Technology, Norway
[2] Department of Materials Science and Engineering, NTNU - Norwegian University of Science and Technology, Norway
* Corresponding author: thomas.tybell@ntnu.no

*Abstract*

Strain engineering with different substrate facets is promising for tuning functional properties of thin film perovskite oxides. By choice of facet, different surface symmetries and chemical bond directions for epitaxial interfaces can be tailored. Here, preparation of well-defined pseudo-cubic (111)-oriented orthorhombic substrates of $DyScO_3$, $GdScO_3$, and $NdGaO_3$ is reported. The choice of orthorhombic facet, $(011)_o$ or $(101)_o$, both corresponding to pseudo-cubic $(111)_{pc}$, gives vicinal surfaces with single or double $(111)_{pc}$ layer terrace step heights, respectively, impacting subsequent thin film growth. Orthorhombic $LaFeO_3$ epitaxy on the $(101)_o$ facet reveals a distinction between alternating $(111)_{pc}$ layers, both during and after growth. The observed differences are explained based on the oxygen octahedral tilt pattern relative to the two orthorhombic $(111)_{pc}$ surfaces. This robust structural detail in the orthorhombic perovskite oxides enables utilisation of different $(111)_{pc}$ facets for property engineering, through polyhedral connectivity control and cation coordination at epitaxial interfaces.

*Introduction*

Perovskite oxides and their interfaces, exhibiting strong structure-property coupling, present a rich playground for controlling fundamental properties including ferroelectricity, (anti)ferromagnetism, and superconductivity [1, 2]. With advances in deposition techniques over the last decades, the synthesis of advanced epitaxial perovskite heterostructures has become possible, exhibiting a wide array of novel and enhanced functionality. Examples include superconductivity at the interface between $SrTiO_3$ (STO) and $LaAlO_3$ (LAO) [3], induced ferroelectricity in a dielectric material by strain [4], and controllable polar skyrmion states in ferroelectric/dielectric superlattices [5]. Moreover, by using different substrate facets for thin film synthesis, functional responses can be tailored. A polar metal has been realised in $LaNiO_3$ on (111)-oriented LAO [6], exchange bias has been engineered at the interface between a ferromagnet and a paramagnet [7], and a switchable ferrimagnetic moment has been demonstrated in an antiferromagnetic material at the interface with a ferromagnetic material [8, 9]. Strain engineering based on (111)-oriented systems has also been predicted by Density Functional Theory (DFT) to stabilise Goldstone-like modes in strained LAO [10] and $SrMnO_3$ [11].

Strain engineering based on non-cubic substrates allows the introduction of anisotropic strain in epitaxial thin films. By relying on $TbScO_3$ and $GdScO_3$ (GSO) substrates, single domain $BiFeO_3$ and $LaFeO_3$ (LFO) have been realised [12, 13], and uniaxial control of the Néel vector has been demonstrated in LFO by strain engineering with pseudo-cubic (111)-oriented orthorhombic substrates [14]. However, for orthorhombic systems there exist two $(111)_{pc}$ facets, $(011)_o$ and $(101)_o$, imposing different strain states in an epitaxial thin film [14]. The subscripts pc, o, and c refer to pseudo-cubic, orthorhombic, and cubic symmetry, respectively. To realise high quality heterostructures, control of substrate surfaces is important and atomically flat substrates with proper surface termination are a prerequisite. For $(001)_{pc}$-oriented rare-earth scandates, it has been reported that $ScO_2$ termination, despite being polar, can be achieved by annealing in $O_2$ flow followed by a basic etch which selectively forms and dissolves rare-earth hydroxides [15]. Moreover, for $(001)_{pc}$ $NdGaO_3$ (NGO), a polar $GaO_2$ termination is achieved by acidic etch prior to annealing in $O_2$ flow [16]. However, little is known about the effect of $(111)_{pc}$-equivalent facets on substrate preparation and subsequent thin film synthesis. In this work, the difference between $(011)_o$ and $(101)_o$ crystalline facets, both corresponding to $(111)_{pc}$, of NGO, $DyScO_3$ (DSO), and GSO substrates is reported. It is shown that the substrate surface morphology depends on the choice of facet, and that thin film synthesis of orthorhombic LFO is different for the two facets, allowing the opportunity for control of octahedral rotations at the interface and surface of orthorhombic $(111)_{pc}$ systems.

*Experimental*

Commercial grade substrates of DSO, GSO, and NGO were bought from Shinkosha (Japan), SurfaceNet (Germany), and PI-KEM (England). The substrates, all orthorhombic (space group 62, *Pbnm*) [17, 18] with $a^-a^-c^+$ Glazer tilt pattern [19], were prepared with a standard miscut, typically ±0.3°, for vicinal surfaces. The surface topography was characterised by atomic force microscopy (AFM, Veeco Nanoscope) in both tapping mode and contact mode, and friction force microscopy (FFM) was used to study the homogeneity of surface preparation. The as-received substrates were found to be flat, but without signs of steps and terraces after the final polishing by the suppliers. Hence, various methods were used to prepare the substrates with atomically flat step-and-terrace surfaces. DSO and GSO were annealed at 1000°C-1050°C for six hours, whereas NGO was etched in a 12.5 % buffered HF/NH$_4$F (BHF) solution for 120 s before annealing at 1000°C for two hours. It is noted that a subsequent basic NaOH etch of the scandates after annealing [15] did not make any difference in the tests performed here. Prior to the surface treatments, all substrates were rinsed in an ultrasonic bath using solvents of acetone, ethanol and de-ionised (DI) water, for five minutes each, before drying under N$_2$ flow before and after the DI water step. For DSO and NGO both (011)$_o$ and (101)$_o$ orientations were used, while for GSO only the (101)$_o$ orientation was used. Thin films of LFO, isostructural with the orthorhombic substrates, were deposited by pulsed laser deposition (PLD) using a KrF excimer laser (248 nm) with a fluence of ∼2 Jcm$^{-1}$ and a repetition rate of 1-3 Hz. The substrates were heated to 540°C during deposition in a $2.5 \times 10^{-3}$ mbar O$_2$ atmosphere, and the substrate target distance was 45 mm [20]. The growth was monitored by reflection high-energy electron diffraction (RHEED) at 25 kV. After deposition the samples were cooled at 15 Kmin$^{-1}$ in a 100 mbar O$_2$ atmosphere. In order to evaluate the surface quality of the LFO thin films, AFM was conducted using an Oxford Instruments/Asylum Research AFM, with tip curvature radius of 10 nm.

*Results and discussion*

AFM topography data for all investigated facets gained after surface treatment is displayed in Figure 1. For each facet a line profile is plotted, denoted by the corresponding white line in the tapping mode AFM scan. In addition, for some facets FFM maps were taken to confirm the homogeneity of the substrate preparation. Figure 1 depicts data for DSO(101)$_o$ (a), DSO(011)$_o$ (b), GSO(101)$_o$ (c), NGO(101)$_o$ (d), and NGO(011)$_o$ (e). All surfaces exhibit clear step-and-terrace features. For the (101)$_o$ facets (a, c, d), the measured height of the steps is 0.47 ± 0.03 nm for DSO (a), 0.49 ± 0.03 nm for GSO (c), and 0.47 ± 0.03 nm for NGO (d). This corresponds to the height of one (101)$_o$ layer, or two (111)$_{pc}$ layers, which are from bulk experimental data 4.48 Å for DSO(101)$_o$, 4.51 Å for GSO(101)$_o$, and 4.44 Å for NGO(101)$_o$ [17, 18]. Double layer steps have been observed consistently over time for all (101)$_o$ facets that have been used, and independent of the substrate being DSO, GSO or NGO. This is in contrast to (001)$_{pc}$-oriented scandate substrates, where reported step heights correspond to single (001)$_{pc}$ layers [21]. For the (011)$_o$ facets (b, e), the steps generally correspond to single (111)$_{pc}$ layers, albeit step bunching is prevalent and can be seen for both DSO (b) and NGO (e). Single step heights are measured to approximately 0.24 ± 0.02 nm for DSO and 0.23 ± 0.02 nm for NGO. These step heights correspond to half the interplanar distance between two (011)$_o$ planes, which are from bulk experimental data 2.32 Å for DSO(022)$_o$ and 2.24 Å for NGO(022)$_o$ [17, 18].

Friction force microscopy was performed on DSO(101)$_o$, GSO(101)$_o$ and NGO(011)$_o$, depicted as inset micrographs in Figure 1(a, c, e). From the FFM scans, overlaid with contact mode AFM scans, the surfaces appear homogeneous with respect to the friction force contrast showing similar levels from all terraces. The layer stacking sequence along [111]$_{pc}$ for DSO, GSO, and NGO is $AO_3^{3-}$ / $B^{3+}$ / $AO_3^{3-}$ / $B^{3+}$ / ..., where A = Dy, Gd, Nd, and B = Sc, Sc, Ga, respectively, resulting in formally ±3 charged polar layers. To stabilise a polar surface and overcome the polar discontinuity, a reconstructed, non-stoichiometric surface layer can form [22, 23]. Surface reconstructions due to polar layer stacking has been observed in (111)$_c$ STO [20, 24], (110)$_c$ STO [25], (110)$_o$ DSO [21], (001)$_{pc}$ LAO [26], and (110)$_{pc}$ LAO [27, 28]. The scanning probe studies presented in Figure 1 thus indicate homogeneously terminated surfaces for both facets, and stable double-layer terrace steps for the (101)$_o$ oriented substrates.

To investigate the effect of the two distinct $(111)_{pc}$ facets on thin film growth, thin films of LFO were synthesised. *In-situ* RHEED data from the specular spot are represented in Figure 2 for LFO grown on DSO$(101)_o$, DSO$(011)_o$, GSO$(101)_o$, and STO$(111)_c$. The LFO/STO$(111)_c$ data is added as a reference for comparison to growth on an ideal cubic substrate. For all facets studied, the first RHEED oscillation period is usually shorter than the following oscillations, which has previously been linked to substrate surface reconstructions when LFO is deposited on $(111)_c$ STO, due to the polarity of the terminating layer [20]. Moreover, a suppression of the RHEED signal is recorded for the initial 4-5 layers of $(111)_{pc}$ LFO growth on the scandates, possibly due to the stabilisation of initial growth and to establish the octahedral rotational structure of the unit cell. At least four $(111)_{pc}$ layers are indeed needed to represent the unit cell of LFO correctly [29] (see also Figure 4), and the signal typically recovers after 5-6 layers. However, the main feature to note is the modulation of the RHEED signal intensity for LFO on $(101)_o$ DSO and GSO. Each RHEED oscillation period corresponds to a single $(111)_{pc}$ layer, as confirmed by thickness fits to XRD data. Hence, the variation in RHEED intensity for every second $(111)_{pc}$ layer indicates a significant difference in scattering from subsequent layers during deposition. The data for LFO on DSO$(011)_o$ does not show periodic modulations of the signal, indicating no difference in scattering between layers. It is noted that these signatures can be readily seen in both the specular and diffractive signals, with electron beam incidence along both in-plane principal axis families $\langle 1\bar{1}0 \rangle_{pc}$ and $\langle 11\bar{2} \rangle_{pc}$. The modulated (non-modulated) RHEED signal has been consistently observed for LFO growth on $(101)_o$ facets ($(011)_o$ facets) also for NGO, and for a large range of oxygen background pressures, $[2.5 \times 10^{-3}, 3.1 \times 10^{-1}]$ mbar, during growth.

To elucidate the observed RHEED modulation of subsequent $(111)_{pc}$ LFO layers on $(101)_o$ facets, *ex-situ* AFM was used. For this purpose, thin LFO films terminated at either a high or low intensity RHEED peak were prepared, see Figure 3(a,b). The RHEED data shows RHEED screen images before and after LFO deposition and the *in-situ* intensity oscillations. The right column shows AFM data for the films. Upper and lower rows correspond to the high and low intensity-terminated films, respectively, as can be seen in the RHEED signal panels. The AFM data shows significant differences in the surface roughness of the individual terraces. On the high intensity-terminated film a root mean square roughness $R_{RMS} = 33 \pm 5$ pm is measured on the terraces (a AFM), whereas $R_{RMS} = 47 \pm 3$ pm is obtained for the low intensity-terminated film (b AFM). The emerging picture for LFO growth on $(101)_o$ and $(011)_o$ orthorhombic perovskites is thus that the two distinct $(111)_{pc}$ layers on the $(101)_o$ facet have different surface roughness resulting in different scattering and an alternating RHEED intensity signal. It is noted that thin film growth on $(101)_o$ facets does not change the step height between terraces, still corresponding to one full $(101)_o$ layer as for $(101)_o$ substrates pre-growth.

In order to rationalise the difference in step heights between the $(101)_o$ and $(011)_o$ facets, a bulk crystal truncation of $(111)_{pc}$ DSO with Sc top layer was considered. The surface layers of a $(101)_o$-oriented and a $(011)_o$-oriented crystal are depicted in Figure 4 (a) and (b), respectively, with surface line scans of the differently oriented substrates for comparison in (c) and (d). Due to the tilt pattern relative to the surfaces, the oxygen octahedra can exhibit two different rotational states, here denoted *u* and *v*, assuming no surface reconstructions. In $(101)_o$ DSO, the angle of the oxygen bonds oriented out-of-plane for *u*-layers form angles of 37°/57° alternating for $a_{pc}/b_{pc}$, and constant 74° for $c_{pc}$ with respect to the surface normal $[111]_{pc}$. The free oxygen bonds of *v*-layers form angles of 60°/76° alternating for $a_{pc}/b_{pc}$, and constant 35° for $c_{pc}$ with respect to the surface normal. In (011)$_o$ DSO the angles are different from the *u*- and *v*-layers for (101)$_o$, however, $a_{pc}$, $b_{pc}$ and $c_{pc}$ alternations instead occur within layers as well as between successive layers. An important difference between the orientations is therefore the stacking sequence of *u* and *v* rotational states, essentially given by the octahedral $c_{pc}$ axis tilt. While the (101)$_o$-orientation has alternating layers of purely *u* and *v* along [111]$_{pc}$, the (011)$_o$-oriented layers along [111]$_{pc}$ each have a mix of *u* and *v*. Assuming that the surface energy is linked to the coordination of the surface oxygen bonds, the surface energy of the layers populated only by *u* or *v* along [111]$_{pc}$ for the (101)$_o$ facet will be different, also with respect to the (011)$_o$ facet. It is noted that a reconstruction of the surface will affect the bond angles and surface energy. However, the surface octahedra will still be restricted by the bonds to the underlying layers, which will either be mixed or pure *u* or *v*.

A similar pattern can be found by considering the A-cation positions instead of the BO$_6$ octahedral tilts, although they are linked, due to the antiferrodistortive shifts of the A-cations. In the case of (101)$_o$ facets, A-cations have alternating shifts within each (101)$_o$ plane that average out, but each plane of A-cations is collectively shifted either up or down along [111]$_{pc}$. This can be seen for subsequent A-cation layers with shifts up/down relative to the B-cations which are located at high-symmetry positions, possibly giving a difference in surface energy between subsequent layers. For the (011)$_o$ facet, the A-cation shifts are alternating both in and out of plane, and hence average out for each subsequent plane giving no difference in surface energy for successive layers.

As the surface oxygen coordination and A-cation positions are structural effects originating from the orthorhombic symmetry with $a^-a^-c^+$ Glazer tilt pattern, these features are readily seen also in GSO, NGO, and orthoferrites, including LFO. In summary, the structural distortions of $BO_6$ octahedra and A-cations in the orthorhombic structure support a difference in surface energy for subsequent $(111)_{pc}$ layers in $(101)_o$ oriented orthorhombic perovskites. Hence, the observed difference in surface roughness and RHEED intensity between the successive $(111)_{pc}$ layers supports a surface energy difference between the *u* and *v* terminations.

*Conclusion*

The stability difference of subsequent $(111)_{pc}$ layers for $(101)_o$ DSO, GSO, and NGO, results in step heights of two $(111)_{pc}$ interplanar distances (~4.5Å) for vicinal surfaces. This is an inherent peculiarity in the orthorhombic perovskite structure, as inferred from experimental data and bulk truncation considerations of the impact of oxygen octahedral bond angles and A-cation distortions. In comparison, the seemingly equivalent facet $(011)_o$ exhibits step heights of one interplanar $(111)_{pc}$ distance (~2.25Å) and no sign of surface energy difference from successive layers, as is normal for (111)-oriented cubic substrates. The difference in surface topography between the two $(111)_{pc}$ facets, and the inferred difference in $(101)_o$ *u/v* surface energies, affect thin film growth of LFO resulting in distinct octahedral rotational terminating layers. The possibility to control cation coordination and polyhedral connectivity across epitaxial interfaces by choice of facet and terminating layer opens new vistas for functional interfaces including engineering of magnetic interactions.

*Acknowledgements*

Zuzana Čiperová is acknowledged for conducting initial surface preparation experiments on DSO and GSO. The Research Council of Norway is acknowledged for providing funding through Grants No. 231290 and 275810, and FRINATEK project no. 263228. Partial funding was obtained from the Norwegian University of Science and Technology (NTNU), and from the Norwegian Ph.D. Network on Nanotechnology for Microsystems, sponsored by the Research Council of Norway, Division for Science, Contract No. 221860/F60.

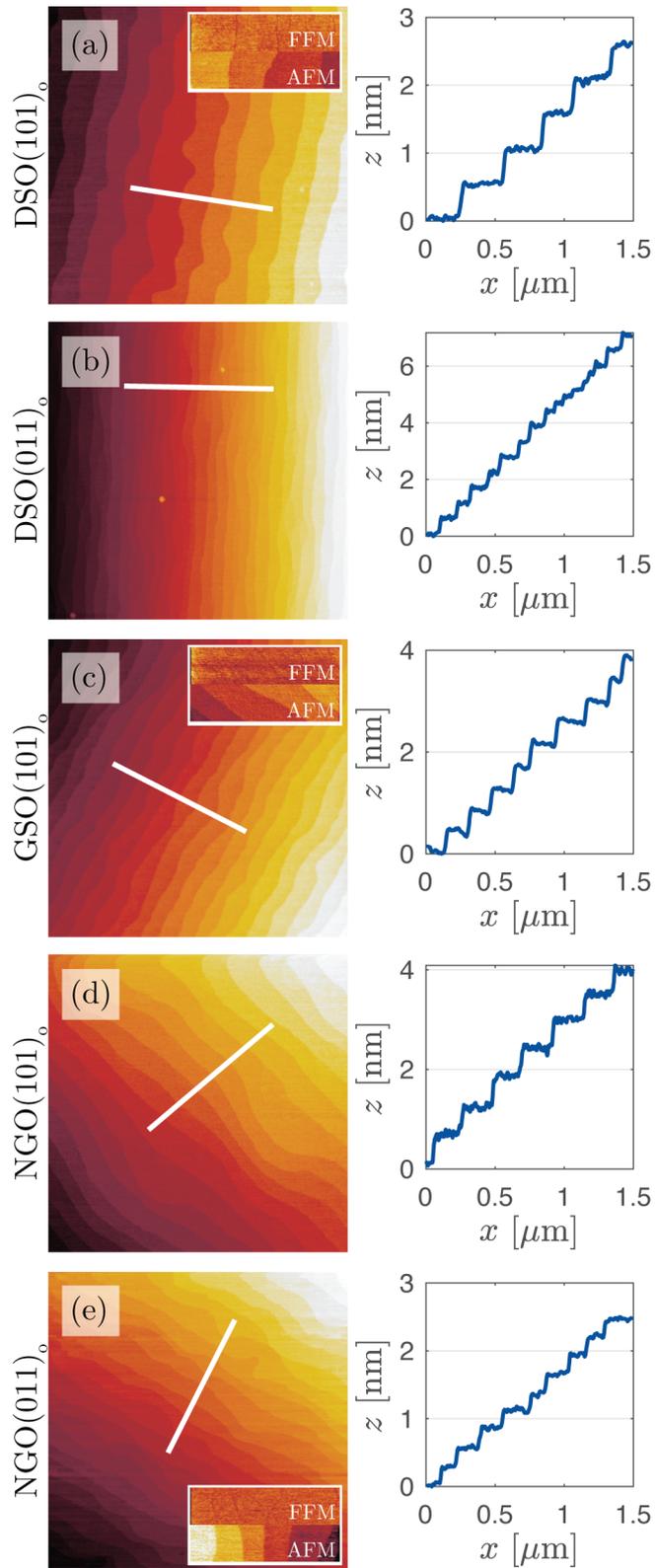

*Figure 1:* Tapping mode AFM micrograph and height profile along indicated white lines for DSO(101)$_o$ (a), DSO(011)$_o$ (b), GSO(101)$_o$ (c), NGO(101)$_o$ (d), and NGO(011)$_o$ (e). In (a), (c) and (e), insets display overlapping contact mode FFM and AFM images. The dimension of all tapping mode AFM micrographs is 3x3 µm², and the contact mode FFM/AFM insets are 1x0.5 µm².

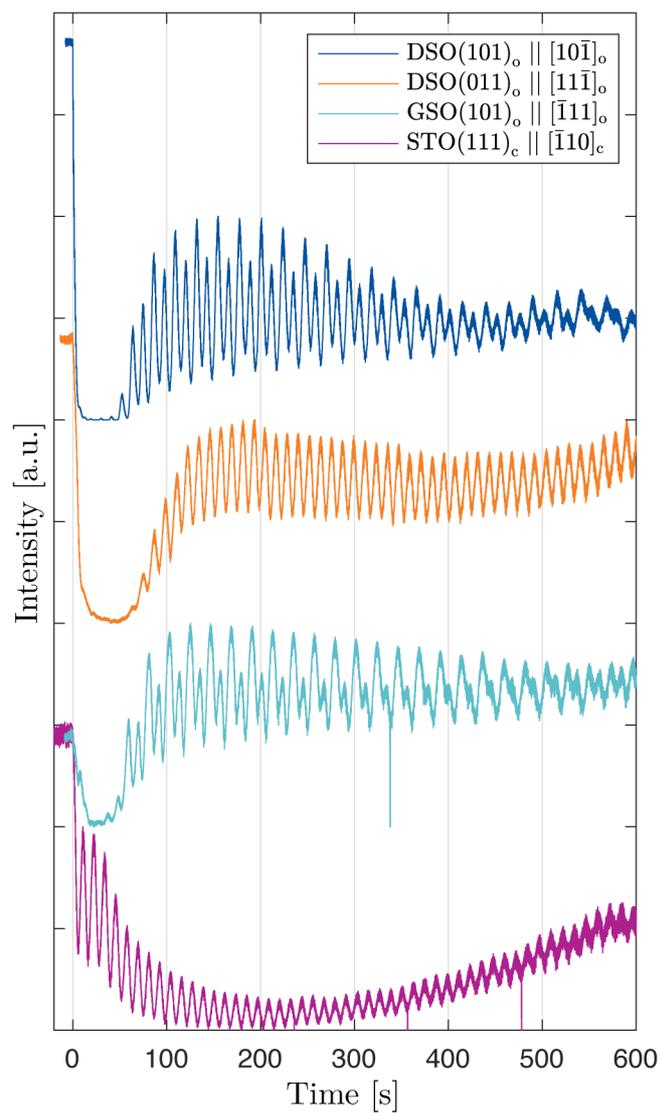

*Figure 2:* Normalised *in-situ* RHEED intensity oscillations for the first 10 minutes of LFO grown on DSO(101)$_o$, DSO(011)$_o$, GSO(101)$_o$, and STO(111)$_c$. Each signal is offset for clarity. Electron beam azimuthal direction is indicated in the legend entries, corresponding to both $\langle 1\bar{1}0 \rangle_{pc}$ and $\langle 11\bar{2} \rangle_{pc}$ directions.

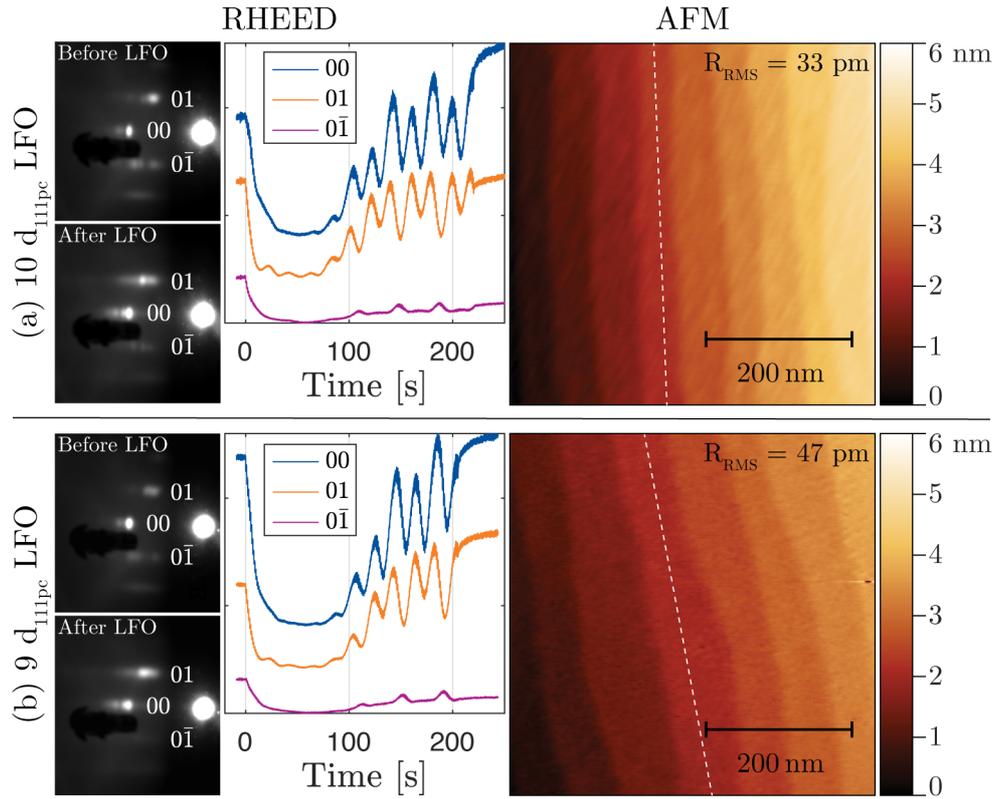

*Figure 3:* RHEED and AFM data for LFO grown on DSO(101)$_o$. The upper row (a) represents 10 d$_{(111)pc}$ LFO layers with growth stopped at a high intensity peak, whereas the lower row (b) corresponds to 9 d$_{111pc}$ LFO layers stopped at a low intensity peak. Columns from left to right show images of the RHEED screen before/after LFO deposition, *in-situ* RHEED intensity for specular and diffractive spots (offset for clarity), and AFM of the LFO films. The step edges seen in the AFM micrographs are parallel to the [10$\bar{1}$]$_o$ direction. Electron beam incidence is along one of the ⟨1$\bar{1}$0⟩$_{pc}$ directions at 30° with respect to the step edges. Indicated RMS roughness values are calculated along the dashed lines and do not vary significantly between terraces.

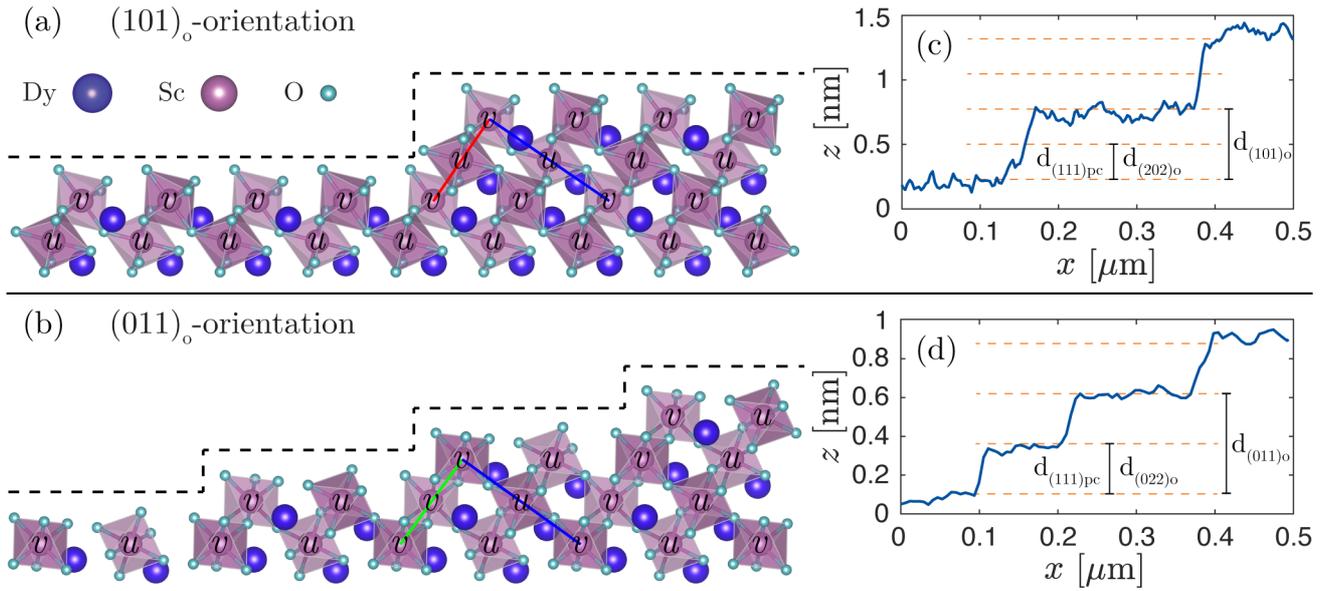

*Figure 4:* Steps and terraces of a (101)$_o$-oriented (a) and (011)$_o$-oriented (b) orthorhombic perovskite, here exemplified by DSO. The *u* and *v* labels indicate different octahedral rotational states with respect to the surface. Red, green and blue lines indicate orthorhombic *a*, *b* and *c* axes, respectively, in *Pbnm* notation. AFM line scans of (101)$_o$-oriented (c) and (011)$_o$-oriented (d) surfaces for comparison. Dashed orange lines indicate (111)$_{pc}$-layers.